\input{aipcheck}

\documentclass[
    ,final            
  ]
  {aipproc}

\layoutstyle{6x9}

\begin{document}

\title{Fit to Electroweak Precision Data\thanks{Presented at the 2006 Conference on the Intersections of Particle and Nuclear Physics (CIPANP 2006), Rio Grande, Puerto Rico, May 30 -- June 3, 2006.}
}

\classification{12.15.-y, 13.66.Jn, 14.80.Bn}
\keywords      {Electroweak interaction; standard model; higgs boson.}

\author{Jens Erler}{address={Instituto de F\'\i sica, Universidad Nacional Aut\'onoma de M\'exico, 04510 M\'exico D.F., M\'exico}}

\begin{abstract}
A brief review of electroweak precision data from LEP, SLC, the Tevatron, and low energies is presented. The global fit to all data including the most recent results on the masses of the top quark and the W boson reinforces the preference for a relatively light Higgs boson.  I will also give an outlook on future developments at the Tevatron Run II, CEBAF, the LHC, and the ILC.
\end{abstract}

\maketitle



\begin{table}[b]
\begin{tabular}{llllr}
\hline
  & \tablehead{1}{r}{b}{central value}
  & \tablehead{1}{r}{b}{uncertainty}
  & \tablehead{1}{r}{b}{SM prediction}
  & \tablehead{1}{r}{b}{pull}   \\
\hline
$R_b$ \hspace*{20pt} &   0.21629 &   0.00066 &   0.21579 &       0.8 \\
$R_c$                          &     0.1721 &     0.0030 &     0.1723 &      -0.1 \\
${\bf A_{FB}^b}$          &\bf 0.0992 &\bf 0.0016 &\bf 0.1031 &\bf  -2.4 \\
$A_{FB}^c$                 &     0.0707 &     0.0035 &     0.0736 &      -0.8 \\
$A_b$                         &     0.923   &     0.020   &     0.9347 &      -0.6 \\
$A_c$                         &     0.670   &     0.027   &     0.6678 &       0.1 \\
\hline
\end{tabular}
\caption{Final LEP and SLC heavy flavor results.}
\label{tab:HF}
\end{table}

\begin{figure}[t]
  \includegraphics[width=330pt]{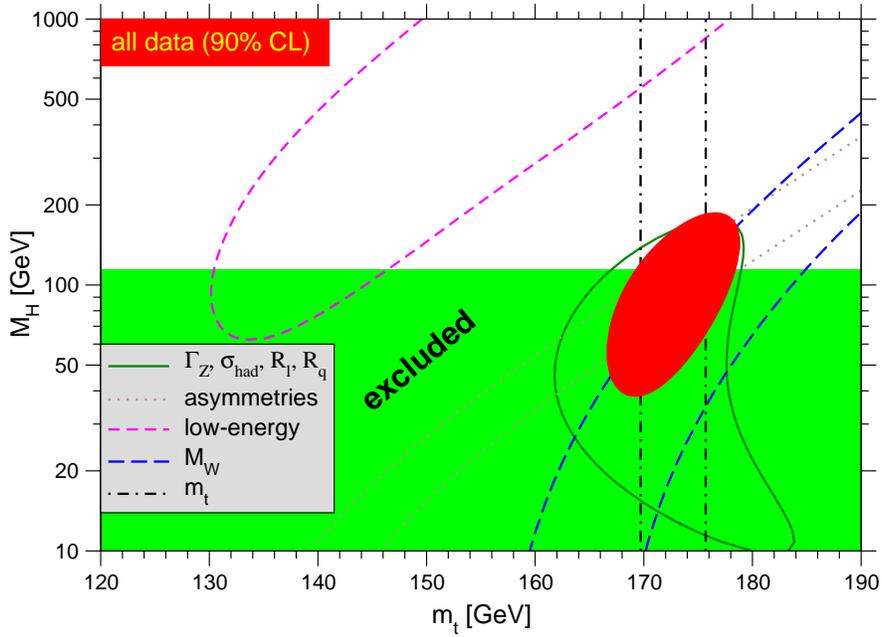}
  \caption{$1\sigma$ constraints and the 90\% CL allowed region by all precision data in the $M_H-m_t$ plane.}
\label{mhmt}
\end{figure}

The Higgs boson remains the only particle of the electroweak Standard Model (SM) which has not been discovered, yet, and constraining its mass, $M_H$, by studying quantum loop effects is currently the prime objective in electroweak physics. Various sets of precision data give complementary constraints on $M_H$ and the top quark mass, $m_t$, and the $M_H-m_t$ plane in Figure~\ref{mhmt} serves as a convenient map of the experimental situation.

The solid (dark green) line is from all $Z$ pole observables~\cite{zpole:2005} other than asymmetries, {\em i.e.\/}, the total $Z$ width, $\Gamma_Z$, the hadronic peak cross section, $\sigma_{\rm had}$, and various partial decay widths normalized to the hadronic $Z$ width. Thus, without reference to any measurement of the weak mixing angle, $\sin^2\theta_W$, or the mass of the $W$ boson, $M_W$, one already recognizes a clear preference for a light Higgs and a top quark mass consistent with the kinematic mass reconstruction by CDF and D\O\ at the Tevatron~\cite{TevatronEWG:2006}.

\begin{table}[b]
\begin{tabular}{lclll}
\hline
  & \tablehead{1}{r}{b}{fb$^{-1}$ per experiment}
  & \tablehead{1}{r}{b}{experimental value}
  & \tablehead{1}{r}{b}{error/{\sl goal}}
  & \tablehead{1}{r}{b}{$\sqrt{L}$-scaling}   \\
\hline
Tevatron Run I         &   0.072 &     0.2238        &\bf 0.0050   &        --- \\
SLC                         &     0.05 &     0.23098      &\bf 0.00026 &        --- \\
LEP~1                     &     0.20 &     0.23187      &\bf 0.00021 &        --- \\
\bf currently              &            &\bf 0.23152      &\bf 0.00016 &        --- \\
Tevatron Run IIA      &     2      &                       &\sl 0.0008    & 0.0009\\
Tevatron Run IIB      &     8      &                       &\sl 0.0003    & 0.0005\\
JLab                         & $\vec{e}e,~\vec{e}p$ & &\sl 0.0003    &       --- \\
LHC high luminosity &  400     &                       &\sl 0.00014 &0.00008\\
ILC                           & M\o ller &                       &\sl 0.00008 &        --- \\
GigaZ                       &   140     &                   &\sl 0.000013&0.000016\\
\hline
\end{tabular}
\caption{Results and future expectations for $\sin^2\theta_W^{\rm eff.}$. Based on $\sqrt{L}$-scaling as appropriate for statistics dominated measurements, the last column extrapolates the Tevatron Run I precision to future hadron colliders. GigaZ, which refers to two years of data taking at a $Z$ factory at the ILC with ${\cal O}(10^9)$ $Z$ bosons, is scaled from the LEP~1 precision. The goal at the LHC is ambitious and assumes almost complete jet rapidity coverage. JLab refers to fixed target scattering at CEBAF where the polarized electron-proton experiment Qweak~\cite{Armstrong:2003gp} is already approved and funded.}
\label{tab:s2w}
\end{table}

The $Z$ pole asymmetries~\cite{zpole:2005} determine the weak mixing angle, $\sin^2\theta_W^{\rm eff.}$. When combined with the Tevatron $m_t$, they give the strongest constraint on $M_H$ (shown as dotted lines) which results from the combination of about a dozen different measurements. But the two most precise ones, the SLD left-right asymmetry, $A_{LR}$, and the LEP forward-backward asymmetry in $b$ quark final states, $A_{LR}^b$, deviate at the $3\sigma$ level from each other. Table~\ref{tab:HF} shows the results from the $Z$ pole heavy flavor sector which were finalized only recently. It shows that $A_{LR}^b$ is significantly lower than the SM expectation.  By contrast, $A_{LR}$ is $2\sigma$ high, and further experimental information on $\sin^2\theta_W^{\rm eff.}$ is urgently needed to help clarify the puzzling situation, which may conceivably even hint at the presence of new physics~\cite{Marciano:CIPANP06}. Table~\ref{tab:s2w} offers a look ahead, where the experimental goals are listed and compared with a simple scaling based on the expected integrated luminosity, $L$.

\begin{figure}[t]
  \includegraphics[width=330pt]{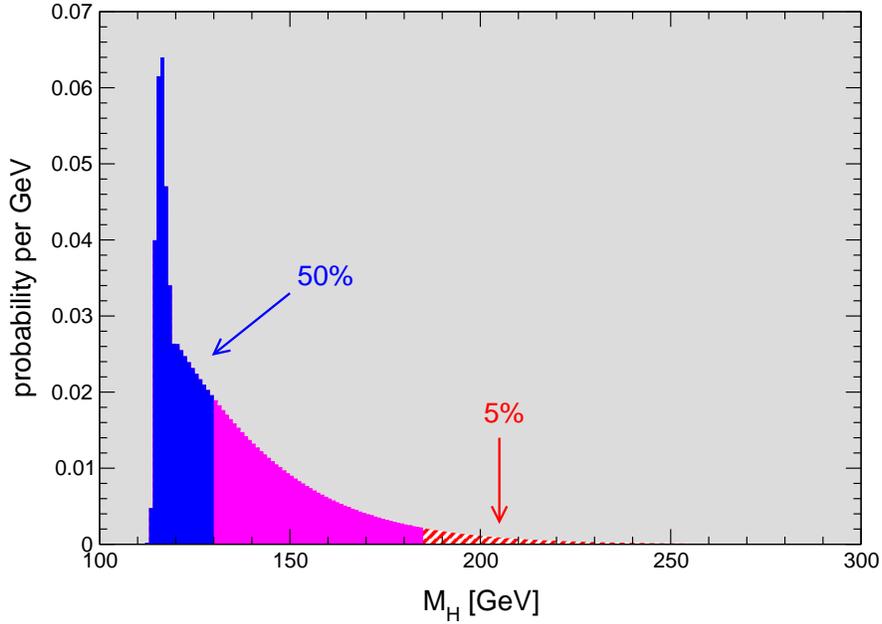}
  \caption{Probability distribution of $M_H$~\cite{Erler:2000cr} based on all precision data and LEP~2~\cite{Barate:2003sz} search results.}
\label{mh}
\end{figure}

The dashed (magenta) contour from measurements at relatively low energies favors higher Higgs masses. This is driven by the NuTeV result~\cite{Zeller:2001hh} on neutrino scattering off left-handed quarks, which shows a $2.7\sigma$ conflict with the SM. There is also a deviation in the muon anomalous magnetic moment muon, which is discussed elsewhere~\cite{Czarnecki:CIPANP06,Miller:CIPANP06}.

\begin{table}[b]
\begin{tabular}{lccrr}
\hline
  & \tablehead{1}{r}{b}{fb$^{-1}$ per experiment}
  & \tablehead{1}{r}{b}{value [GeV]}
  & \tablehead{1}{r}{b}{error/{\sl goal}}
  & \tablehead{1}{r}{b}{$\sqrt[4]{L}$-scaling}   \\
\hline
Tevatron Run I         &     0.11 &      80.452 &\bf 59 &     --- \\
LEP 2                       &     0.71 &     80.388 &\bf 35 &     37 \\
\bf currently              &\bf 0.81 &\bf 80.405 &\bf 30 &\bf 36 \\
Tevatron Run IIA      &     2      &                 &\sl 31 &     29 \\
Tevatron Run IIB      &     8      &                 &\sl 25 &     20 \\
LHC low luminosity  &    10     &                 &\sl 23 &     19 \\
LHC high luminosity &  400     &                 &\sl  9 &        8 \\
ILC                           &  300     &                 &\sl 10 &       8 \\
MegaW                    &    70     &                &\sl 7 &4$^*$\hspace*{-4pt}\\
\hline
\end{tabular}
\caption{Results and future expectations for $M_W$. The last column extrapolates the Tevatron Run I precision under the assumption that sensitivities scale as in background dominated types of experiments. The exception is MegaW which refers to a dedicated threshold scan at the ILC with ${\cal O}(10^6)$ $W$ pairs and is based on a $\sqrt{L}$-scaling from a similar scan at LEP~2. As can be seen, $\sqrt[4]{L}$-scaling provides a simple estimate of future precision goals in this case.}
\label{tab:MW}
\end{table}

Constraints from $M_W$ (long-dashed) are becoming increasingly competitive with those from $\sin^2\theta_W^{\rm eff.}$. LEP~2 data are still being analyzed and future prospects are summarized in Table~\ref{tab:MW}. Similarly, $m_t$ results (dot-dashed) and error projections are shown in Table~\ref{tab:mt}.

\begin{table}[t]
\begin{tabular}{lccrr}
\hline
  & \tablehead{1}{r}{b}{fb$^{-1}$ per experiment}
  & \tablehead{1}{r}{b}{value [GeV]}
  & \tablehead{1}{r}{b}{error/{\sl goal}}
  & \tablehead{1}{r}{b}{$\sqrt[4]{L}$-scaling}   \\
\hline
Tevatron Run I         &     0.11 &      178.0 &\bf 4.3 &     ---  \\
summer 2005          &     0.43 &      172.7 &\bf 2.9 &     3.1 \\
\bf currently              &\bf 0.86 &\bf 172.5 &\bf 2.3 &\bf 2.6 \\
Tevatron Run IIA      &     2      &               &\sl 2.0 &      2.1 \\
Tevatron Run IIB      &     8      &               &\sl 1.2 &      1.5 \\
LHC low luminosity  &    10     &               &\sl 0.9 &      1.4 \\
LHC high luminosity &  400     &               &\sl 0.7 &      0.6 \\
ILC                           &  300     &               &\sl 0.1 &       --- \\
\hline
\end{tabular}
\caption{Results and future expectations for $m_t$. The last column extrapolates the Tevatron Run I precision assuming that sensitivities scale as in background dominated types of experiments. At hadron colliders, a $\pm 0.6$~GeV theory uncertainty has to be added, because the kinematic mass determined there is a long-distance mass presumably close to the pole mass. The conversion from the pole mass to a short-distance mass (like $\overline{\rm MS}$) which actually enters the electroweak loop corrections is plagued by an irreducible uncertainty of the order of the strong interaction scale~\cite{Smith:1996xz}. At the ILC it would be possible to determine the $\overline{\rm MS}$ mass directly.}
\label{tab:mt}
\end{table}

Combining all precision data, we find $M_H = 88^{+34}_{-26}$~GeV and $m_t = 172.5 \pm 2.3$~GeV. For the strong coupling constant we obtain, $\alpha_s (M_Z) = 0.1216 \pm 0.0017$, where we also use the $\tau$ lifetime and leptonic branching ratios as constraints. The minimum $\chi^2$ is 47.4 for 42 effective degrees of freedom. The probability for a larger $\chi^2$ is 26\%. Omitting the top quark mass from the Tevatron, yields $m_t = 172.2^{+10.0}_{\phantom{0}-7.4}$~GeV in excellent agreement with CDF and D\O. The 90\% range based only on precision data, $47\mbox{ GeV} < M_H < 146$~GeV, is to be compared with the 95\% upper limit, $M_H \leq 185$~GeV, which also takes the results on LEP~2 Higgs boson searches~\cite{Barate:2003sz} into account (the LEP~2 search limit is $M_H > 114.4$~GeV). The $M_H$ probability distribution is shown in Figure~\ref{mh}.


\begin{theacknowledgments}
It is a pleasure to thank Paul Langacker for collaboration and the organizers of CIPANP 2006 for a very enjoyable meeting. This work was supported by CONACyT (M\'exico) contract 42026--F and by DGAPA--UNAM contract PAPIIT IN112902.
\end{theacknowledgments}



\bibliographystyle{aipproc}   

\bibliography{sample}

\IfFileExists{\jobname.bbl}{}
 {\typeout{}
  \typeout{******************************************}
  \typeout{** Please run "bibtex \jobname" to optain}
  \typeout{** the bibliography and then re-run LaTeX}
  \typeout{** twice to fix the references!}
  \typeout{******************************************}
  \typeout{}
 }


\end{document}